# Sparse geometries handling in lattice Boltzmann method implementation for graphic processors

Tadeusz Tomczak, Roman G. Szafran

**Abstract**—We describe a high-performance implementation of the lattice Boltzmann method (LBM) for sparse geometries on graphic processors. In our implementation we cover the whole geometry with a uniform mesh of small tiles and carry out calculations for each tile independently with proper data synchronization at the tile edges. For this method, we provide both a theoretical analysis of complexity and the results for real implementations involving two-dimensional (2D) and three-dimensional (3D) geometries. Based on the theoretical model, we show that tiles offer significantly smaller bandwidth overheads than solutions based on indirect addressing. For 2D lattice arrangements, a reduction in memory usage is also possible, although at the cost of diminished performance. We achieved a performance of 682 MLUPS on GTX Titan (72% of peak theoretical memory bandwidth) for the D3Q19 lattice arrangement and double-precision data.

**Index Terms**—GPU, CUDA, LBM, CFD, parallel computing

## 1 INTRODUCTION

The LBM is a highly parallel alternative to classical Navier-Stokes solvers for computational fluid dynamics (CFD). Many researchers have confirmed its excellent absolute performance and scalability on modern parallel machines, especially graphic processing units (GPUs). A thorough review of GPU LBM implementations is presented in [1].

In many areas of CFD application (e.g., biomedical or porous media simulations), the geometry is sparse (i.e., it contains many areas without fluid). Although many optimization techniques for LBM implementations in GPUs are known [1]–[11], they are not specialized for sparse geometries. Simulations for sparse geometries can be run using methods designed for large-scale multi-GPU implementations [12]–[14], but the lack of specific optimizations can significantly increase memory usage and computational effort required for simulation, which decreases the domain size that can be simulated on a single machine and increases the number of processors required for a domain of a given size. This can be especially unfavorable for GPUs, where the memory amount per processor is fixed and usually smaller than in typical CPU configurations. Thus, the efficient handling of sparse geometries is crucial in order to achieve high performance and good hardware utilization.

Current GPU implementations of the LBM, specialized for sparse geometries, as presented in [15]–[19], use indirect addressing. This limits memory usage but decreases performance. For example, highly optimized sparse implementation from [18] achieves 337 MLUPS, whereas implementa-

- *Tadeusz Tomczak is the Chair of Computer Engineering at Wroclaw University of Science and Technology, Faculty of Electronics, ul. Janiszewskiego 11/17, 50-370 Wroclaw, Poland.*
  *Email: tadeusz.tomczak@pwr.edu.pl*
- *Roman G. Szafran is based at Wroclaw University of Science and Technology, Faculty of Chemistry, Department of Chemical Engineering, ul. Norwida 4/6, 50-373 Wroclaw, Poland.*
  *Email:roman.szafran@pwr.edu.pl, phone: +48713203813.*

tion for dense geometries from [10] reaches 526 MLUPS for the same collision model (BGK quasi-compressible, D3Q19, double-precision) and hardware (Tesla K20; although, in [10], ECC was disabled, giving an additional 10%).

In this work, we present the GPU implementation of the LBM for sparse geometries where the information about geometry is stored by using the uniform grid of small, fixed-size tiles. The use of tiles reduces the values of a number of ancillary data transfers, which are needed to manage the geometry sparsity, to almost negligible values. For this method, we present a theoretical analysis, which allows us to find the performance limits and compare them with indirect addressing solutions. We also show the results for two real implementations: a 2D implementation with minimized memory usage and a 3D implementation with high performance.

The structure of the paper is as follows. In Section 2, we briefly introduce the LBM and analyze existing techniques of sparse geometries handling. Section 3 describes our implementation with a detailed analysis of the introduced overheads. The performance comparison with existing implementations is presented in Section 4. Section 5 contains conclusions.

## 2 LATTICE BOLTZMANN METHOD

### 2.1 Basics

In the LBM, as in conventional CFD, the geometry, initial and boundary conditions must be specified to solve the initial value problem. The computational domain is uniformly partitioned, with computational nodes placed in vertices of adjacent cells, which become the lattice. One of the lattice structures used in this study, D2Q9, is presented in Fig. 1. The first number in the notation (D2) is the space dimension, while the second (Q9) is the lattice links number.

Let $f_i$ represent the probability distribution function (PDF) along the $i$ lattice direction: in which case, $\delta_x$ and $\delta_t$ are the lattice spacing and the lattice time step, respectively;

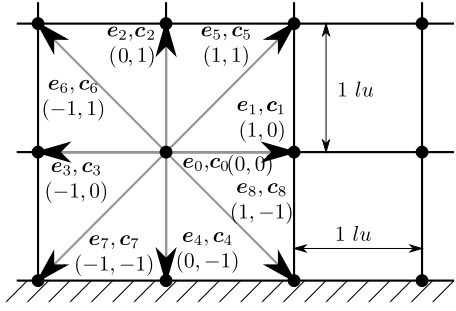

Fig. 1. The D2Q9 lattice structure; $e_i$, $c_i$ are vectors along the $i$ lattice direction; $lu$ denotes the lattice unit.

$\frac{\delta_x}{\delta_t}$ is the lattice speed, $e_i$ is the unit vector along the $i$ lattice direction, and $c_i = \frac{\delta_x}{\delta_t} e_i$ is the lattice velocity vector in the velocity space along the $i$ lattice direction. The core of the LBM is discretized in velocity $R^n$ space form of the Boltzmann transport equation, which can be written for each $i$ of the $q$ directions (lattice linkages) in the velocity space as follows:

$$\frac{\partial f_i}{\partial t} + c_i \cdot \nabla f_i = \Omega_i, \quad (1)$$

where $\Omega_i$ is the collision operator.

The collision operator can be approximated with the most popular Bhatnagar-Gross-Krook (BGK) model,

$$\Omega_{BGK} = -\frac{1}{\tau}(f_i - f_i^{eq}) \quad (2)$$

where $\tau$ is the relaxation time in lattice Boltzmann (LB) units related to the lattice fluid viscosity: $v = 1/3(\tau - 1/2)$. $f_i^{eq}$ is the equilibrium distribution function along the $i$ lattice direction given by the following formula, in the case of the quasi-compressible model:

$$f_i^{eq} = \omega_i \rho \left( 1 + \frac{c_i \cdot u}{c_s^2} + \frac{(c_i \cdot u)^2}{2c_s^4} - \frac{u^2}{2c_s^2} \right) \quad (3)$$

and as follows in the case of the incompressible model:

$$f_i^{eq} = \omega_i \left( \rho + \frac{c_i \cdot u}{c_s^2} + \frac{(c_i \cdot u)^2}{2c_s^4} - \frac{u^2}{2c_s^2} \right) \quad (4)$$

where $c_s = 1/\sqrt{3}$ is the lattice speed of sound, which is a lattice constant; $u$ is the macroscopic fluid velocity vector expressed in LB units; $\rho = \sum_i f_i$ is a fluid density, which is related to a pressure $p = \rho/3$, both expressed in LB units; and $\omega_i$ is a weighting scalar for the $i$ lattice direction. The macroscopic velocity for the quasi-compressible model can be determined by:

$$u = \frac{1}{\rho} \sum_i c_i f_i \quad (5)$$

and as follows for the incompressible model:

$$u = \sum_i c_i f_i. \quad (6)$$

By integrating Eqn. (1) from $t$ to $t + \delta_t$ along the $i$ lattice direction, and assuming that the collision term is constant during the interval, we can obtain a form of the BGK-LBM equation discretized in time:

$$\underbrace{f_i(r + c_i \delta_t, t + \delta_t) - f_i(r, t)}_{Streaming} = \underbrace{\frac{\delta_t}{\tau}[f_i^{eq}(r, t) - f_i(r, t)]}_{Collision}, \quad (7)$$

where $r$ is a position vector in the velocity space. The therm on the left-hand side is known as the streaming (propagation) step, while the latter represents the collision step. These two steps are repeated sequentially during the simulation, providing velocity, density and pressure distributions at each time step.

The BGK-LBM uses the single relaxation time to characterize the collision effects. However, physically, these rates should be different during collision processes. To overcome this limitation, a collision matrix with different eigenvalues or multiple relaxation times can be used. The LBM with a multiple relaxation time (MRT) collision operator can be expressed as:

$$\underbrace{f_i(r + c_i \delta_t, t + \delta_t) - f_i(r, t)}_{Streaming} = \underbrace{A[f^{eq} - f]}_{Collision}, \quad (8)$$

where $A$ is the collision matrix. In this study, we use both of the above-mentioned collision models in compressible and quasi-compressible versions.

### 2.2 Performance model

To compare different implementations, we employ a widely used, simple computational complexity model based on [20]. In this model, we assume that data from the device memory are only read once and cached in the fast internal memory (registers/cache/scratchpad), and that the device memory can be effectively read in single-byte transactions. These assumptions are rather unrealistic for current machines because both CPUs and GPUs use DRAM memories with burst transactions and have registers and a cache of limited size. However, such an "ideal" machine model allows us to find minimal amounts of data that need to be stored and transferred during computations. The minimal amounts of data define both the minimum memory usage and the maximum achievable performance for bandwidth-bound implementations.

For simplification, we use the same measure of complexity for fluid and boundary nodes. In many geometries, the boundary nodes are only a small portion of all non-solid nodes. Moreover, many boundary nodes do not require significantly different amounts of stored and transferred data. Even if additional values are used, they are often shared between many nodes and can be placed in the cache/constant memory.

Let $q$ denote the number of PDFs $f_i$ and $s_d$ denote the number of bytes for storing a single $f_i$ value (e.g., 4/8 B for a single-/double-precision floating point). Thus, the minimum number of memory bytes needed for a single node datum is:

$$M_{node} = q \cdot s_d \quad [B] \quad (9)$$

and the minimum number of bytes transferred per single LBM iteration for a single node is:

$$B_{node} = 2 \cdot q \cdot s_d \quad [B]. \quad (10)$$

Notice that we assume that information about the node type is not stored in the memory.

Formulas defining the number of floating-point operations (FLOP, not to be confused with FLOPS) are much more complex to determine. A simple count of the operations resulting from a naive implementation of equations (7) and (8) gives numbers that are significantly larger than in real implementations, due to optimizations carried out at the compilation stage (multiplications by constants $e_i \in \{-1, 0, 1\}$, common subexpression elimination, constant folding etc.). Thus, for numbers of computational operations, we show only specific values obtained by disassembling a GPU binary code using the *nvdisasm* utility (we only count floating-point arithmetic operations, as the fused multiply-add operation is treated as two operations). The complete number of FLOP per D2Q9 lattice node (including computations of velocity and density) varies from 52 FLOP for the BGK incompressible model to 145 FLOP for the MRT quasi-compressible model, where the floating-point inverse must be computed on a GPU. For the D3Q19 lattice, the computations require between 304 FLOP for the BGK incompressible model and 1,165 FLOP for the MRT quasi-compressible model.

The numbers of FLOP cannot be treated as the minimal numbers of operations, but can be used to estimate whether a performance of LBM implementation on a given machine is bound by memory bandwidth, instruction throughput or latency. According to Eqn. (10), each node requires the transfer of 144 bytes for D2Q9 and 304 bytes for D3Q19 lattice arrangements and double-precision data. This gives between 2.77 B/FLOP (BGK incompressible model) and 0.99 B/FLOP (MRT quasi-compressible model) for the D2Q9 lattice arrangement, and between 1 B/FLOP (BGK incompressible model) and 0.26 B/FLOP (MRT quasi-compressible model) for the D3Q19 lattice arrangement.

Comparing these numbers with those defined in [21] for the *machine balance* of GPUs (about 0.29 B/FLOP for Fermi, 0.18 B/FLOP for Kepler, and 0.14 B/FLOP for Pascal architectures), it can be seen that LBM implementations within GPUs should usually be bandwidth-bound. Only the performance of MRT quasi-compressible implementation for the D3Q19 lattice arrangement on Fermi architectures can be limited by arithmetic operations.

## 2.3 Sparsity handling overhead

Compared with the minimal requirements defined in Section 2.2, the techniques for sparse geometries result in some memory and bandwidth overheads. We ignore the computational overhead because the LBM implementations within GPUs are usually bandwidth-bound. Let the overhead $\Delta$ ($\Delta^M$ - memory overhead, $\Delta^B$ - bandwidth overhead) be defined as a ratio of additional operations to the minimum numbers defined in Eqn. (9) and (10). The performance may be then estimated as $1/(1 + \Delta^B)$, assuming that the performance of the ideal implementation equals 1. Only a single LBM time iteration is analyzed, since all iterations are processed in the same way.

Let $N_{snodes}$ and $N_{fnodes}$ denote the number of solid and non-solid (fluid and boundary) nodes in a geometry. The number of all nodes in the geometry is then $N_{nodes} = N_{snodes} + N_{fnodes}$. We can also define a geometry porosity as follows:

$$\phi = \frac{N_{fnodes}}{N_{nodes}} \quad (11)$$

and a solidity thus:

$$\eta = \frac{N_{snodes}}{N_{nodes}}, \quad (12)$$

where both $\phi, \eta \in [0, 1]$ and $\phi + \eta = 1$.

### 2.3.1 Connectivity matrix

The GPU implementations for a connectivity matrix (CM), shown in [15] and as an optimized version in [18], use pointers to the neighbor node for each propagated function $f_i$. No data are stored for solid nodes, but the additional pointers for each $f_i$ function are needed for each non-solid node. Since the pointer can be replaced by an index to a linear array, for GPU implementations, the $s_{CM\_idx} = 4$ byte index is enough (6 GB allows for storing less than $200 \cdot 10^6$ nodes for the D2Q9 lattice and a single copy of single-precision $f_i$ values). Thus, each non-solid node requires an additional storage of $q - 1$ indices ($q$ for implementation according to [18]), while the memory overhead is:

$$\Delta^M_{CM} = \frac{(q-1) \cdot s_{CM\_idx}}{M_{node}} + 1, \quad (13)$$

where +1 denotes the second copy of $f_i$ data for each non-solid node used to avoid race conditions in both [15] and [18]. For double-precision $f_i$ values and $s_{CM\_idx} = 4$ bytes, the left term in Eqn. (13) is less than 0.5.

The bandwidth overhead for CM only results from the readings of the indices from the CM. Thus, using the minimum transfer for a single node, as defined in Eqn. (10), the bandwidth overhead is:

$$\Delta^B_{CM} = \frac{(q-1) \cdot s_{CM\_idx}}{B_{node}} \approx \frac{s_{CM\_idx}}{2 \cdot s_d}. \quad (14)$$

The value of $\Delta^B_{CM}$ depends mainly on the ratio of $s_{CM\_idx}$ to $s_d$, and is $< 0.5$ for single-precision and $< 0.25$ for double-precision $f_i$ values.

### 2.3.2 Fluid index array

The GPU implementation of fluid index array (FIA) technique, presented in [19], uses a "bitmap", which, for each node, contains either a pointer to the non-solid node data or −1 when a node is solid. The values of $f_i$ functions are stored only for non-solid nodes. Thus, the additional memory is used only for pointers from the FIA (single pointer per solid/non-solid node). Similar to the CM, pointers can be replaced with $s_{FIA\_idx} = 4$ B indices. The memory overhead for the FIA is:

$$\Delta^M_{FIA} = \frac{s_{FIA\_idx} \cdot N_{nodes}}{N_{fnodes} \cdot M_{node}} + 1 = \frac{s_{FIA\_idx}}{\phi \cdot M_{node}} + 1, \quad (15)$$

where +1 denotes the second copy of $f_i$ data for each non-solid node used in [19] to avoid race conditions.

For geometries with a low number of solid nodes, the left term from Eqn. (15) is much smaller than 1 (about $\frac{0.111}{\phi}$ for single-precision and about $\frac{0.056}{\phi}$ for double-precision $f_i$ values and the D2Q9 lattice arrangement). However, when a geometry is very sparse, the value of $\frac{s_{FIA\_idx}}{\phi \cdot M_{node}}$ grows fast: for the D2Q9 lattice arrangement and single-precision $f_i$ values, it exceeds 1.0 for $\phi = 0.1$.

The bandwidth overhead for FIA-based implementation is more difficult to define. Fast LBM implementations fuse collision and streaming steps into a single kernel. Unfortunately, the original implementation from [19] uses two separate kernels. The first kernel, launched only for fluid nodes, reads and writes $f_i$ values and realizes the collision step. The second kernel, launched for all nodes (also the solid ones), is responsible for the streaming step; in addition, it reads and writes the $f_i$ values, as well as reads the FIA indices for the current node and for all neighbor nodes, if the current node is non-solid. This approach significantly increases the bandwidth overhead (to values larger than 1) due to double access to $f_i$ values.

In our estimation, we treat the additional reads/writes of $f_i$ values as being necessary for FIA-based implementations and model these by adding the "+1" term to the bandwidth overhead. The remaining overhead only results from additional reads of indices from the FIA. Since we assume that the indices from the FIA are only read from the device memory once and cached, the bandwidth overhead is:

$$\Delta^B_{FIA} = \frac{N_{nodes} \cdot s_{FIA\_idx}}{N_{fnodes} \cdot B_{node}} + 1 = \frac{s_{FIA\_idx}}{\phi \cdot B_{node}} + 1. \quad (16)$$

Notice that $\frac{s_{FIA\_idx}}{\phi \cdot B_{node}}$ is equal to half of the left term from Eqn. (15).

### 2.3.3 Other methods

Simulations for sparse geometries can be also run using codes for dense geometries, usually on multi-GPU clusters due to large memory requirements. Depending on the applied solutions, the memory and performance penalties caused by geometry sparsity can vary.

For the simplest case, all data must be stored and transferred for all nodes (fluid and solid), thus $1/\phi$ times more memory and transfers are needed. For example, storing data for all nodes from a geometry with $\phi = 0.1$ requires $10\times$ more memory than when only storing data for fluid nodes. Since an FIA/CM require about two to $2.5\times$ more memory per node than implementations for dense geometries, the use of code designed for dense geometries to simulate the sparse ones requires about $4-5\times$ more memory when $\phi = 0.1$.

The bandwidth overhead of implementation for dense geometries can be greatly reduced by skipping operations for solid nodes after checking for the type of node. This optimization can reduce the bandwidth overhead to very low values. However, this does not take into account the uncoalesced memory transactions resulting from interlacing in memory data for solid and non-solid nodes.

The interesting technique is presented in waLBerla, the large-scale multi-GPU framework [12]–[14], where the geometry is divided using the hierarchical structure of "patches" composed of "blocks". For sparse geometries, empty blocks can be removed, thus reducing memory usage and computational complexity.

Although tiles presented in this work use a similar concept, blocks involve a higher-level data structure designed for efficient multiprocessor implementations, where load balancing and communication affect performance.

The main differences between tiles and blocks are size (tiles have a small, fixed size and are additionally limited by hardware parameters, for example, warp/memory transaction/cache line size, whereas blocks are much larger, up to more than $100^3$ nodes, in order to minimize communication overhead), additional data (blocks contain management information), data layout (because of small tile size, a special memory layout minimizing the number of uncoalesced transactions is needed), and hardware mapping granularity (blocks are mapped per single process, whereas tiles are processed by separate GPU thread blocks).

Tiles presented in this work may be considered as the optimized implementation of multilevel hierarchical grids, with only a single level of resolution and without transformations between different levels of resolution. Hierarchical grids, as proposed in [22], are used in many LBM simulators designed for multi-CPU machines: HemeLB [23], [24], Palabos [25], OpenLB [26] and Musubi [27].

## 3 TILES

In our implementation, the whole geometry is covered by the uniform mesh of fixed-size tiles. Let $a$ denote the number of nodes per tile edge. The number of nodes per tile is $n_{tn} = a^2$ for 2D and $n_{tn} = a^3$ for 3D lattice arrangements. If a geometry size is not divisible by $a$, the geometry is extended with solid nodes.

The tiling is implemented by the host code and performed once at the geometry load. We use a very simple algorithm: first, the geometry is covered by the uniform mesh of tiles, starting at node (0,0), after which the tiles containing only solid nodes are removed.

During a single LBM time step, the tiles can be processed independently and in any order, provided that values at the tile edges are correctly propagated. We implemented two methods for data when synchronizing the tile edges.

The first method, that is, tiles with two copies (T2C), uses two copies of $f_i$ data and the *gather* pattern, e.g., values from neighbor nodes are accessed during the read stage. The information about tile placement is stored in an additional *tileMap* array with pointers to data for non-empty tiles (for empty tiles, $-1$ is used). The pointers are used to find the neighbor tiles needed during the propagation stage. The *tileMap* array is similar to the FIA, but uses tile granularity, which decreases overheads proportionally to the number of nodes per tile.

In the second method, that is, tiles with ghost buffers (TGB), we use only one copy of $f_i$ data and additional *ghost buffers* at the tile edges. The ghost buffers contain copies of propagated $f_i$ functions. During propagation in a single LBM time step, the values from ghost buffers can be read and written at the same time (and in any order). To prevent data races from occurring, we use two copies of ghost buffers: one copy is used for the data store, while the other copy is used for the data read. After each LBM time iteration, the copies are exchanged (by pointers exchange). The idea behind two-step propagation is depicted in Fig. 2. For data inside the tiles, we use the *scatter* pattern, where the computed values are stored in neighbor nodes. For ghost buffers, the *gather* pattern is used.

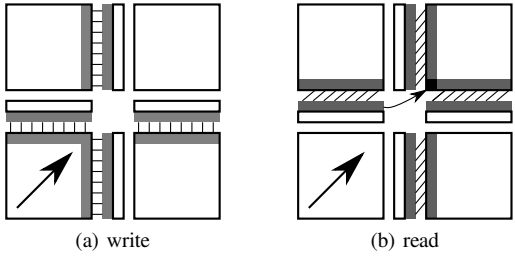

Fig. 2. Two-step propagation toward the north-east (see arrow) of values for edge nodes in tiles with ghost buffers. Nodes and ghost buffers used in operations are marked as gray. Notice that edge values are shifted during data reads (writes are done without shifting). In this way, we avoid uncoalesced writes for values at the corners; only reads may be uncoalesced (black node in the right image).

## 3.1 Tiles overhead

Compared with the minimal requirements defined in Section 2.2, the tiles cause some memory, bandwidth and computational overheads due to solid nodes inside tiles, additional data for storing information about tile placements, additional data for storing node type, methods of avoiding race conditions (either two copies of $f_i$ data or ghost buffers), uncoalesced memory transactions due to data layout and the computational overhead for address calculations, among others. The reasons for overheads fall into one of the two classes: the overheads caused by method and the overheads resulting from implementation.

In this section, we focus exclusively on the former class. For simplification, we consider only those lattice arrangements that require just a single layer of ghost buffers (up to D2Q9/D3Q19 for 2D/3D geometries).

Let $n_{tsn}$ denote the average number of solid nodes per tile, and let $n_{tfn} = n_{tn} - n_{tsn}$ denote the average number of non-solid nodes per tile. We can define the average tile porosity as follows:

$$\phi_t = \frac{n_{tfn}}{n_{tn}} = \frac{n_{tn} - n_{tsn}}{n_{tn}} = 1 - \frac{n_{tsn}}{n_{tn}} = 1 - \eta_t, \quad (17)$$

where $\eta_t$ is the average tile solidity, $\eta_t, \phi_t \in [0, 1]$ and $\eta_t + \phi_t = 1$. The minimum memory required to store all non-solid nodes from a tile is then:

$$M_{tile} = n_{tfn} \cdot M_{node} = n_{tn} \cdot \phi_t \cdot M_{node} \quad [B] \quad (18)$$

and the minimum transfer for a tile is:

$$B_{tile} = n_{tfn} \cdot B_{node} = n_{tn} \cdot \phi_t \cdot B_{node} \quad [B]. \quad (19)$$

### 3.1.1 Memory overhead

The memory overhead for tiles is a sum of four components: the overhead resulting from solid nodes inside a tile $\Delta^M_{\phi_t}$, the memory required to store node type $\Delta^M_{nt}$, the memory used to avoid race conditions $\Delta^M_{rc}$, and the memory needed for storing the additional tile data $\Delta^M_{ad}$.

For all nodes in a tile (independently of the node type), all $f_i$ values must be stored. Thus the memory overhead resulting from solid nodes in a tile is equal to:

$$\Delta^M_{\phi_t} = \frac{n_{tsn} \cdot M_{node}}{M_{tile}} = \frac{n_{tsn}}{n_{tfn}} = \frac{1}{\phi_t} - 1. \quad (20)$$

Let $s_t$ denote a number of bytes used to store the node type (we assume that at least one byte is necessary, despite the fact that, for simple cases, a few bits may be enough). Since the node type is stored for all nodes inside a tile, the overhead from the memory required for node type values is:

$$\Delta^M_{nt} = \frac{n_{tn} \cdot s_t}{M_{tile}} = \frac{s_t}{M_{node}} \cdot \frac{1}{\phi_t}. \quad (21)$$

The values of $\Delta^M_{rc}$ and $\Delta^M_{ad}$ depend on the method of race condition prevention.

3.1.1.1 Two copies: In this method, an additional copy of all $f_i$ values is needed for each node in a tile. The memory overhead is then:

$$\Delta^M_{rc} = \frac{n_{tn} \cdot M_{node}}{M_{tile}} = \frac{1}{\phi_t}. \quad (22)$$

The additional tile datum is the *tileMap* array where one index per tile (empty and non-empty) is stored. In practice, the index is $s_{ti} = 4$ bytes in width; this allows us to use $2^{32}$ non-empty tiles, which require much more memory than is available in current GPUs.

Let $N_{tiles}$ denote the number of all tiles and $N_{ftiles}$ denote the number of non-empty tiles. The overhead resulting from the memory needed to store the *tileMap* array is then:

$$\Delta^M_{ad} = \frac{N_{tiles} \cdot s_{ti}}{N_{ftiles} \cdot M_{tile}} = \frac{N_{tiles} \cdot s_{ti}}{N_{ftiles} \cdot \phi_t \cdot n_{tn} \cdot M_{node}}. \quad (23)$$

Notice that, since $s_{ti}/(n_{tn} \cdot M_{node}) < 0.01$, even for the D2Q9 lattice, single-precision data and small tiles containing $n_{tn} = 4^2$ nodes, $\Delta^M_{ad}$ is negligible unless the ratio $N_{tiles}/N_{ftiles}$ is large. For geometries used in this work, $N_{tiles}/N_{ftiles} \in (2.3, 8.6)$.

The complete memory overhead for two copies of $f_i$ data is then:

$$\Delta^M_{T2C} = \frac{1}{\phi_t} \cdot \left(2 - \phi_t + \frac{1}{M_{node}} \cdot \left(s_t + \frac{N_{tiles}}{N_{ftiles}} \cdot \frac{s_{ti}}{n_{tn}}\right)\right). \quad (24)$$

Usually the components $\Delta^M_{nt}$ and $\Delta^M_{ad}$ can be skipped ($1/M_{node} < 0.03$, even for D2Q9 lattice and single-precision data), which allows us to approximate the memory overhead as $\Delta^M_{T2C} \approx (2 - \phi_t)/\phi_t$.

Equation (24) contains many parameters; but, for concrete implementation, most of them are constant. For example, for double-precision data ($s_d = 8$), for two bytes per node type ($s_t = 2$) and for $s_{ti} = 4$, the memory overhead is:

$$\Delta^M_{T2C} = \frac{2.028 + 0.00022 \cdot \frac{N_{tiles}}{N_{ftiles}}}{\phi_t} - 1 \quad (25)$$

where the D2Q9 lattice and tiles containing $16^2$ nodes are involved. For the D3Q19 lattice and $4^3$ nodes per tile:

$$\Delta^M_{T2C} = \frac{2.013 + 0.00041 \cdot \frac{N_{tiles}}{N_{ftiles}}}{\phi_t} - 1. \quad (26)$$

3.1.1.2 Ghost buffers: The memory needed for ghost buffers at tile edges depends on two factors. The first is the number of $f_i$ functions propagated in directions toward the faces, the edges and the corners. The second is the number of common faces, edges and corners between neighbor tiles. For example, even if the current tile has only neighbors contacting either the vertexes (for 2D/3D) or the edges (for 3D), the additional buffers on the edges/faces must be allocated (see the black node in the corner of Fig. 2b). Where tiles are contacting vertexes, only single-element ghost buffers can be allocated; however, to simplify the implementation, we assume that ghost buffers are allocated only in full sizes ($n_{tn}/a$ values per buffer, which gives $a$ for 2D and $a^2$ for 3D geometries). The performance penalty caused by this simplification is small: for test cases, as shown in Table 1 in Section 4, less than 2% of allocated ghost buffers have the reduced size.

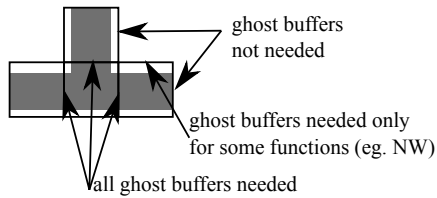

Fig. 3. If some tiles are removed, the ghost buffers are only needed for the edges between non-empty tiles. Non-solid nodes are marked as gray.

In practice, some ghost buffers are unnecessary. Even for geometries that are completely filled with tiles, the ghost buffers at the geometry edges are not used. If the geometry is sparse and some tiles are removed, the corresponding ghost buffers can be removed as well (Fig. 3). To include these removed ghost buffers, we use an additional geometry-dependent coefficient, $\alpha_M$, which is the ratio of allocated ghost buffers to all possible ghost buffers. In real cases, $\alpha_M < 1$ because $\alpha_M = 1$ only applies to infinite geometries without edges. For our test cases, $\alpha_M > 0.7$ was separately computed for each case after the geometry tiling (see Table 1).

To prevent data races from occurring, each PDF $f_i$ needs a set of two ghost buffers: one for a read and one for a write. Let $q_s$ denote the number of $f_i$ functions for which only one set of ghost buffers is needed (the functions propagated toward the edges in 2D and toward the faces in 3D geometries), $q_d$ denote the number of $f_i$ functions for which two sets of ghost buffers are needed (see Fig. 2, where the functions are propagated toward the corners in 2D and toward the edges in 3D geometries), and let $q_t$ denote the number of functions requiring three sets of ghost buffers (the functions propagated toward the corners for 3D geometries). For the D2Q9 lattice, we get $q_s = 4, q_d = 4, q_t = 0$; for the D3Q19 lattice, the values are $q_s = 6, q_d = 12, q_t = 0$; and, for D3Q27, $q_s = 6, q_d = 12, q_t = 8$. The number $q$ of all $f_i$ functions is then $q = q_s + q_d + q_t + 1$ (+1 represents the non-propagated function $f_0$). The maximum number of ghost buffers per tile is $2 \cdot (q_s + 2 \cdot q_d + 3 \cdot q_t)$.

By having these relationships, the average (per tile) real memory usage for ghost buffers is:

$$M_{gb} = 2 \cdot (q_s + 2 \cdot q_d + 3 \cdot q_t) \cdot \frac{n_{tn}}{a} \cdot s_d \cdot \alpha_M \quad [B] \quad (27)$$

where $s_d$ is defined in Section 2.2 as the number of bytes for storing a single $f_i$ value. By using $M_{tile}$, as defined in Eqn. (18), the memory overhead resulting from the ghost buffers is:

$$\Delta_{rc}^M = \frac{M_{gb}}{M_{tile}} = C_{gb} \cdot \frac{2 \cdot \alpha_M}{a \cdot \phi_t} \quad (28)$$

where $C_{gb} = (q_s + 2 \cdot q_d + 3 \cdot q_t)/q$ is a constant defined for a given lattice arrangement. For D2Q9, D3Q19 and D3Q27 lattice arrangements, the value of $C_{gb}$ is $\frac{4}{3}$, $\frac{30}{19}$ and 2, respectively.

The last element affecting the memory overhead is the memory needed for the additional tile data. Most of this memory comprises pointers to ghost buffers. Since pointers can be replaced with array indices, in typical cases, each pointer (index) requires $s_{gbi} = 4$ bytes, which allow us to address more ghost buffers than can be stored in current generation GPUs.

Due to the *gather* pattern (data propagation during the read stage), for each tile, we need to use $q_s + 2 \cdot q_d + 3 \cdot q_t$ pointers to ghost buffers that are written, and $q_s + 3 \cdot q_d + 7 \cdot q_t$ pointers to ghost buffers that are read. The overhead resulting from the storage of ghost buffer indices is then:

$$\Delta_{ad}^M = C_{gbi} \cdot \frac{s_{gbi}}{M_{tile}} \quad (29)$$

where $C_{gbi} = 2 \cdot q_s + 5 \cdot q_d + 10 \cdot q_t$ is a constant defined for a given lattice arrangement. For the D2Q9, D3Q19 and D3Q27 lattice arrangements, there are, respectively, $C_{gbi} = \{28, 72, 152\}$ indices per tile.

The complete memory overhead for tiles with ghost buffers is then:

$$\Delta_{TGB}^M = \frac{1}{\phi_t} \cdot \left(1 - \phi_t + \frac{1}{M_{node}} \cdot \left(s_t + \frac{C_{gbi} \cdot s_{gbi}}{n_{tn}}\right) + \frac{2 \cdot \alpha_M \cdot C_{gb}}{a}\right). \quad (30)$$

For double-precision data ($s_d = 8$), where there are two bytes per node type ($s_t = 2$ and $s_{gbi} = 4$) the memory overhead is:

$$\Delta_{TGB}^M = \frac{1.034 + 0.167 \cdot \alpha_M}{\phi_t} - 1 \quad (31)$$

where the D2Q9 lattice and tiles containing $16^2$ nodes are involved. For the D3Q19 lattice and $4^3$ nodes per tile:

$$\Delta_{TGB}^M = \frac{1.043 + 0.789 \cdot \alpha_M}{\phi_t} - 1. \quad (32)$$

Notice that, for the D3Q19 lattice arrangement, the memory overhead $\Delta_{rc}^M$ caused by ghost buffers is almost $5\times$ larger than for D2Q9, mainly due to $4\times$ smaller tile edge.

### 3.1.2 Bandwidth overhead

In the bandwidth overhead estimation, we assume that, for each node in each non-empty tile, only the node type field must be read ($s_t$ bytes per node). When a node is solid, no other operations are done; thus no unnecessary transfers of $f_i$ values occur. The $f_i$ values for non-solid nodes are also transferred from/to either the tile or ghost buffers, such that, for each node, only a minimal amount of $f_i$ data (see Eqn. (10)) is read/written. In this way, the bandwidth overhead only results from transfers of the node type field $\Delta_{nt}^B$ and transfers of additional tile data $\Delta_{ad}^B$.

For simplification, we also ignore the fact that some writes of $f_i$ values can be skipped when the neighbor node is solid. Additionally, since we use the same measure for fluid and boundary nodes, each non-solid node requires the same amount of transferred data.

The node type field must be read, not only for the current node, but also for all $(q-1)$ neighbor nodes of every non-solid node in order to avoid propagation to/from solid nodes. However, we assume that the node type fields for a single tile are internally buffered in some way (registers/cache/scratchpad); thus, for each tile, the node type fields are only read once for nodes inside a tile and once for nodes forming a "halo" (one node in width) around the tile. The value of $\Delta_{nt}^B$ is then:

$$\Delta_{nt}^B = \frac{(a+2)^d \cdot s_t}{B_{tile}} \qquad (33)$$

where $d \in \{2,3\}$ is the space dimension and $a$ is the number of nodes per tile edge. The value of $\Delta_{ad}^B$ depends on the method of race conditions prevention.

3.1.2.1 Two copies: In this method, each non-empty tile needs to load pointers to the neighbor tiles required for propagation; thus, $q-1$ indices of $s_{ti}$ bytes are loaded for each non-empty tile. We do not need to load the index of the current tile because it can be computed in another way. The bandwidth overhead caused by additional data is then:

$$\Delta_{ad}^B = \frac{(q-1) \cdot s_{ti}}{B_{tile}} \qquad (34)$$

and the complete bandwidth overhead for the T2C method is:

$$\Delta_{T2C}^B = \frac{1}{B_{tile}} \cdot \left((a+2)^d \cdot s_t + (q-1) \cdot s_{ti}\right). \qquad (35)$$

For double-precision data, two bytes per node type and four bytes per tile index $s_{ti}$, the bandwidth overhead $\Delta_{T2C}^B = \frac{0.0184}{\phi_t}$ for the D2Q9 lattice with $16^2$ nodes per tile, and the bandwidth overhead $\Delta_{T2C}^B = \frac{0.0259}{\phi_t}$ for the D3Q19 lattice and $4^3$ nodes per tile.

3.1.2.2 Ghost buffers: For implementation based on ghost buffers, each non-empty tile needs to load $C_{gbi}$ indices of $s_{gbi}$ bytes each. Although some indices can be skipped when all nodes placed on the tile face/edge are solid, the detection of such a situation seems to involve a complex operation. Moreover, the number of skipped indices depends on geometry; in fact, this number may be proportional to $(1-\alpha_M)$, which, for cases shown in Table 1, is smaller than 0.3. Fig. 9 also shows that the value of $\Delta_{ad}^B$ is significantly lower than $\Delta_{nt}^B$ for all test geometries; thus, the decrease of $\Delta_{ad}^B$, even by 30%, results in a much lower decrease in the total overhead. Therefore, in the below considerations, we can safely assume that each non-empty tile always loads all indices onto the ghost buffers. The bandwidth overhead caused by additional data is then:

$$\Delta_{ad}^B = \frac{C_{gbi} \cdot s_{gbi}}{B_{tile}} \qquad (36)$$

and the complete overhead for the TGB method is:

$$\Delta_{TGB}^B = \frac{1}{B_{tile}} \cdot \left((a+2)^d \cdot s_t + C_{gbi} \cdot s_{gbi}\right). \qquad (37)$$

For double-precision data, two bytes per node type and four bytes per the ghost buffer index $s_{gbi}$, the bandwidth overhead $\Delta_{T2C}^B = \frac{0.0206}{\phi_t}$ for the D2Q9 lattice with $16^2$ nodes per tile, and the bandwidth overhead $\Delta_{T2C}^B = \frac{0.0370}{\phi_t}$ for the D3Q19 lattice and $4^3$ nodes per tile.

3.1.2.3 Burst transactions impact: The overhead estimates presented above define the minimal possible overheads obtainable on an ideal machine (Section 2.2). For real parallel machines, additional overheads appear.

Modern machines with DRAM memory use *burst* transactions, including an aligned block of at least $s_b$ bytes (usually $s_b \in \{32, 64, 128\}$). As such, several neighbor values are always transferred during a single memory transaction. To achieve the minimal bandwidth overhead, the node data must be placed in the memory in a way that allows us to fully utilize all memory transactions. Since this may be difficult, we also need to consider the bandwidth overhead for implementation with a suboptimal memory layout.

Due to the spatial and temporal locality of data references, it seems reasonable to assume that the maximum transfer per tile contains, in addition to the node type fields and the additional tile data, all $f_i$ values from tile and all $f_i$ values from ghost buffers (for tiles with ghost buffers). We should also take into account that, for ghost buffers containing the values from tile corners, only a single transaction is needed to read a single $f_i$ value (writes of single corner values do not occur).

Notice that, for 3D tiles, we assume that, even for ghost buffers containing only a single edge of values ($a$ values), the full ghost buffer with $a^2$ values is transferred because it does not require special data arrangements.

The bandwidth overhead resulting from transfers of full tile data is:

$$\Delta_{ftd}^B = \frac{n_{tn} \cdot B_{node} - B_{tile}}{B_{tile}} = \frac{1}{\phi_t} - 1. \qquad (38)$$

Let $q_c$ denote the number of corner values that are read in separate $s_b$ byte transactions ($q_c = q_d$ for 2D and $q_c = q_t$ for 3D lattice arrangements). The additional bandwidth resulting from transfers of all ghost buffers contains two components, that is, the transfers of all ghost buffers, except the ones containing only a single corner value:

$$B_{gbnc} = (C_{gbi} - q_c) \cdot \frac{n_{tn}}{a} \cdot s_d, \qquad (39)$$

and the reads of all corner values:

$$B_{gbc} = q_c \cdot s_b. \qquad (40)$$

In real geometries, some ghost buffers are not needed; thus, the maximum transfer should be scaled by the coefficient $\alpha_B \leq 1$ (similar to $\alpha_M$), which is defined as the ratio of actually transferred ghost buffer values to the maximum number of transferred ghost buffer values. Notice that $\alpha_B$ slightly differs from $\alpha_M$ because, for tiles, the read ghost buffers differ from the written ghost buffers. In our sparse 2D geometries, $\alpha_B$ is slightly lower than $\alpha_M$ ($\alpha_B > 0.9 \cdot \alpha_M$). The bandwidth overhead, including the burst transactions impact, is thus:

$$\Delta_{T2Cbt}^B = \Delta_{T2C}^B + \Delta_{ftd}^B \qquad (41)$$

where there are tiles with two copies; and:

$$\Delta_{TGBbt}^B = \Delta_{TGB}^B + \Delta_{ftd}^B + \frac{B_{gbnc} + B_{gbc}}{B_{tile}} \cdot \alpha_B \qquad (42)$$

where there are tiles with ghost buffers. Both values defined in Eqn. (41) and Eqn. (42) may be treated as pessimistic estimates of the bandwidth overhead.

```
1:  load node type
2:  if node not solid then
3:      load all f_i
4:  end if
5:  load ghost buffers {2 BARRIERS due to WLP}
6:  if node not solid then
7:      process boundary nodes {calculate u, v, ρ}
8:      if node is boundary then
9:          store all f_i
10:     end if
11:     collide
12: end if
13: store ghost buffers {1 BARRIER due to WLP}
14: if node not solid then
15:     scatter all f_i
16: end if
```

Fig. 4. Structure of the GPU kernel implementing a single LBM time iteration for a single node within the tile with ghost buffers containing $16^2$ nodes. WLP denotes warp-level programming.

```
1:  load copy of tile bitmap {3^3 values}
2:  load node types from current tile {4^3 values}
3:  load node types from neighbor tiles {WLP}
4:  BARRIER
5:  if node not solid then
6:      load f_0
7:      for i ∈< 1..18 > do
8:          compute address of neighbor node in direction i
9:          if neighbor node not solid then
10:             gather f_i from neighbor node
11:         end if
12:     end for{all f_i copied to registers}
13:     process boundary {calculate v, ρ}
14:     collide
15:     store all f_i {all coalesced}
16: end if
```

Fig. 5. Structure of the GPU kernel implementing a single LBM time iteration for a single node within the tile containing $4^3$ nodes with two copies of $f_i$ data.

## 3.2 Implementation details

The code was written in the CUDA C Version 7.5 programming language, the first version with official support for many C++11 features. The support for C++11 allowed us to design a heavily templated generic kernel code, which can be specialized for fluid and collision models, tile size and enabled optimizations.

In our implementation, a single GPU thread block processes a single tile, which allows for the easy synchronization of computations within the tile. Consecutive GPU threads are assigned to nodes from tiles using row order. To minimize the number of memory transfers, we combine collision, propagation (streaming) and boundary computations for a single node into a single GPU kernel. The node data ($f_i$, $v$, $\rho$, node type) are transferred only once, with local copies stored in registers and the shared memory.

Figs. 4 and 5 show the two versions of the kernel. We start from the version of the TGB method for the D2Q9 lattice arrangement and $16^2$ nodes per tile, after which we prepare the version that uses the T2C method for the D3Q19 lattice arrangement and $4^3$ nodes per tile, where we also fix issues observed in the D2Q9 version. The main difference between the TGB and the T2C implementations is in the propagation step.

The propagation between two nodes is only performed when the target and source nodes are not solid. In both implementations, we are able to use the node type fields for nodes inside the tile, but nodes from neighbor tiles have to be treated in a different way. For the TGB method, partial information about the node type for nodes from neighbor tiles is stored in the values put into the ghost buffers: if the node corresponding to the given value in the ghost buffer is solid, the value in the ghost buffer is set as the predefined "not a number". In the T2C version, the node type fields from neighbor tiles are tested directly.

In both versions, we use shared memory to store either the ghost buffers that are propagated horizontally (line 5 in Fig. 4) or node types from current and neighbor tiles (Lines 2-3 in Fig. 5). This allows us to minimize the memory bandwidth usage caused by multiple reads of the same values and uncoalesced memory access. However, since the number of either ghost buffers or node type fields for current tiles with an additional halo does not map well onto GPU threads, we use WLP to balance the load of warps assigned to a tile. Although WLP requires additional synchronization in the kernel, it is amortized with interest: in the kernels without WLP, both the performance and the real GPU utilization are lower. The unfavorable issue with the WLP-based solution is the low code portability.

Besides the method of providing information about node types from neighbor tiles, the second difference between the TGB and the T2C implementations of propagation is the data access pattern. In the TGB version, the propagation is implemented by scattering data to neighbor nodes and additionally partitioned into two separated steps; this is because the propagation of $f_i$ values for nodes at the tile edges requires a sequence of writes and subsequent reads to/from ghost buffers (see Fig. 2).

Notice that both these propagation steps must be done before the collision, and that synchronization between the first and second steps is required to prevent data races for nodes at the tile edges (because the tiles are processed in parallel). Thus, our single iteration starts with the second step of propagation (fifth line in Fig. 4) and performs the first propagation step at the iteration end (Lines 13-16).

Additionally, for the propagation performed as data scattering, some $f_i$ values cannot be correctly updated by the neighbor nodes (for example, when these nodes are solid). To avoid this, Lines 8-10 in Fig. 4 are responsible for writing the values for these $f_i$ functions, which will not be overwritten during propagation, to the global memory.

For the T2C kernel, we implement the propagation by gathering data from neighbor nodes. This allows us to simplify the code because no equivalent of Lines 8-10 from Fig. 4 is needed. Due to a lack of access to ghost buffers, we can also remove synchronization points during writes (all writes are only to nodes within a tile and are perfectly coalesced), which allows for a much simpler, regular code with the single barrier at Line 4.

The last important difference between the TGB and the T2C kernels is in the neighbor node address computations (Line 8 in Fig. 5, not shown in Fig. 4). In the TGB version, the address for neighbor node data only depends on the location of the current node and can be determined according to current tile data. The T2C version additionally needs the locations of neighbor tiles; thus, the two-step procedure is needed. First, the index of the tile containing the neighbor node is computed: we use the values from $\{-1, 0, 1\}$ due to the local copy of the tile bitmap (as created in Line 1). Then, when the neighbor tile is not empty, a few indices for the neighbor node are computed, since we need information about the neighbor node type (this is obtained from the copy in the shared memory) and the value of the proper $f_i$ (which is stored in the global memory).

We also apply a set of widely known optimizations: loop unrolling to increase instruction-level parallelism, due to the interlacing of independent streams of instructions; using the shared memory to decrease register pressure; tuning the usage of registers for each specialization of the kernels; minimizing the cost of divergent branches by placing time-consuming memory access outside the divergent code; replacing divisions with multiplications by inversion; avoiding transfers of $v, \rho$ values; and using in-line methods and compiler pragmas to connect clean, structured code with high performance.

Additionally, for the T2C kernel, we use an optimized data layout in the memory for each array of $f_i$ values in order to minimize the number of uncoalesced memory transactions. After this optimization, the real number of memory transactions for each completely filled tile is very close to the minimum value defined by Eqn. (19), due to increased cache utilization. The address calculations are performed by using *constexpr* methods.

## 4 RESULTS

All simulations were run in double precision on a computer with the NVIDIA GTX TITAN device (Kepler architecture), clocked at 823 MHz with a 6 GB GDDR5, 3.004 GHz memory, an Intel i7-3930K CPU and a 64 GB four-channel DDR3 DRAM. Simulations for the D3Q19 lattice arrangement were done using the T2C kernel, while the TGB kernel was used for D2Q9. The tile size was $16^2$ nodes for 2D and $4^3$ nodes for 3D geometries.

### 4.1 Test geometries

The simulations were run for dense (fluid flow in a square chamber with a moving lid) and sparse 2D and 3D geometries. The sparse geometries for the D3Q19 lattice arrangement were prepared on the basis of similar cases, as presented in [18] (three arrays of randomly arranged spheres with a diameter of 40 lattice units and blood flow in a cerebral aneurysm) and [19] (aorta with coarctation). We also performed some tests on arrays with randomly arranged spheres with porosities lower than those used in [18]. For the D2Q9 lattice arrangement, we used the microvascular structures from [28] (Fig. 6 ChipA) and [29] (ChipB). The parameters of the cases are shown in Table 1. The ratio of the boundary to all non-solid nodes varied from 0.05 (for the cerebral aneurysm) up to 0.5 (for RAS_0.1). The porosity of geometries varied from 0.09 to 0.9.

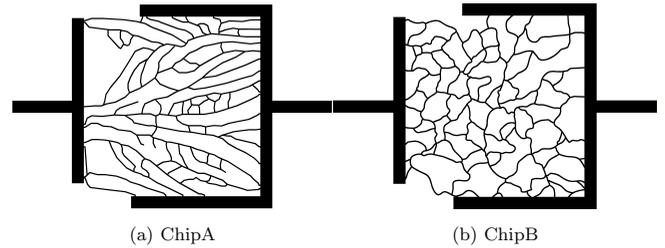

(a) ChipA    (b) ChipB

Fig. 6. Geometries for 2D test cases. Non-solid nodes are marked as black. Channel width is defined for narrow channels in the middle of the drawing.

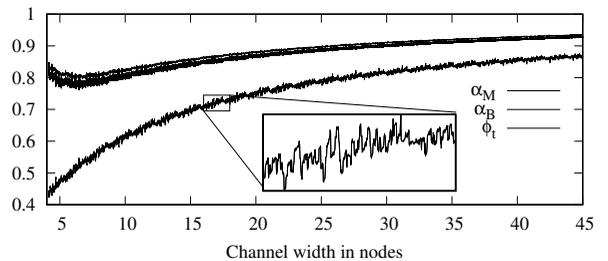

Fig. 7. Tile parameters for ChipA geometry.

#### 4.1.1 Tile parameters

The value of the most important tile parameter, tile porosity $\phi_t$, mainly depends on factors other than geometry porosity $\phi$. This is especially visible in 2D cases, where different values of $\phi_t$ are obtained for the same $\phi$. Fig. 7 shows tile parameters as a function of the number of nodes per channel width. Only ChipA is analyzed; however, the values are similar for ChipB. The value of $\phi_t$ greatly depends on the channel width: $\phi_t$ becomes larger than $0.8$ for 27 node-wide channels ($1.7\times$ the tile edge) and larger than $0.85$ for 39 node-wide channels ($2.4\times$ the tile edge). An increase of $\phi_t$, resulting from a larger channel diameter, can also be observed in 3D geometries: the channel diameter for the aneurysm ($\phi_t = 0.93$) and coarctation ($\phi_t = 0.81$) cases are about 60 and 30 nodes ($15\times$ and $7.5\times$ the tile edge), respectively. The values of $\alpha_M$ and $\alpha_B$ are between $0.76$ and $0.97$. Both coefficients are close to each other (the difference was below $0.073$), while $\alpha_B$ is usually lower than $\alpha_M$.

#### 4.1.2 Memory overhead

The comparison of memory overhead estimates for sparse test geometries is shown in Table 1 and Fig. 8. Fig. 8 also contains the components of both $\Delta_{T2C}^M$ and $\Delta_{TGB}^M$. The results show that, when tile porosity $\phi_t$ is high ($\phi_t > 0.5 \ldots 0.86$), tiles offer the lowest memory usage. For low $\phi_t$, indirect addressing methods require less memory.

In all presented cases, the memory overhead for the TGB method is lower than for the CM and FIA methods. Additional analysis of RAS geometries with porosities lower than $0.7$ showed that, according to Eqn. (32), the CM results in lower memory usage for 3D geometries with tile porosity $\phi_t < 0.7$ ($\phi < 0.3$ for RAS cases). For all 2D geometries, the memory usage for the TGB method is lower than for the CM, even for $\phi_t$ as low as $0.58$ (ChipA_08 case). According to Eqn. (31), for 2D geometries, the TGB method should result in less memory usage than for the CM when $\phi_t > 0.5$.

TABLE 1
Cases used for tests: $N_{nodes}$ - the number of all nodes; $\phi$ - geometry porosity as defined by Eqn. (11); $\phi_t$ - average tile porosity as defined by Eqn. (17); $\Delta^M$, $\Delta^B$ - memory and bandwidth overheads for the T2C, TGB, CM and FIA methods; and $\alpha_M$ - the coefficient as defined in Section 3.1.1.2. The suffixes after the *ChipA*/*ChipB* names denote the channel width in the nodes. The suffixes after the randomly arranged sphere (*RAS*) names denote porosity.

| Case | $N_{nodes}$ | $\phi$ | $\phi_t$ | $\alpha_M$ | $\Delta^M$ | | | | $\Delta^B$ | | | |
|---|---|---|---|---|---|---|---|---|---|---|---|---|
| | | | | | TGB | T2C | FIA | CM | TGB | T2C | FIA | CM |
| RAS_0.9 | 7 077 888 | 0.90 | 0.97 | 0.97 | **0.86** | 1.08 | 1.03 | 1.47 | 0.038 | **0.027** | 1.015 | 0.24 |
| RAS_0.8 | 7 077 888 | 0.80 | 0.94 | 0.96 | **0.92** | 1.15 | 1.03 | 1.47 | 0.040 | **0.028** | 1.016 | 0.24 |
| RAS_0.7 | 7 077 888 | 0.70 | 0.90 | 0.94 | **0.99** | 1.24 | 1.04 | 1.47 | 0.041 | **0.029** | 1.019 | 0.24 |
| Aneurysm | 84 607 488 | 0.18 | 0.93 | 0.97 | **0.95** | 1.17 | 1.15 | 1.47 | 0.040 | **0.028** | 1.075 | 0.24 |
| Coarctation | 6 990 336 | 0.09 | 0.81 | 0.91 | **1.19** | 1.50 | 1.28 | 1.47 | 0.046 | **0.032** | 1.140 | 0.24 |
| ChipA_08 | 2 392 350 | 0.21 | 0.58 | 0.80 | **1.01** | 2.49 | 1.27 | 1.44 | 0.035 | **0.032** | 1.133 | 0.22 |
| ChipB_08 | 2 587 417 | 0.20 | 0.60 | 0.82 | **0.94** | 2.37 | 1.27 | 1.44 | 0.034 | **0.031** | 1.137 | 0.22 |
| ChipA_16 | 9 574 320 | 0.20 | 0.71 | 0.86 | **0.65** | 1.85 | 1.27 | 1.44 | 0.029 | **0.026** | 1.137 | 0.22 |
| ChipB_16 | 10 370 560 | 0.20 | 0.74 | 0.87 | **0.58** | 1.73 | 1.28 | 1.44 | 0.028 | **0.025** | 1.141 | 0.22 |
| ChipA_32 | 38 075 466 | 0.20 | 0.83 | 0.91 | **0.43** | 1.45 | 1.28 | 1.44 | 0.025 | **0.022** | 1.139 | 0.22 |
| ChipB_32 | 41 455 800 | 0.20 | 0.84 | 0.92 | **0.42** | 1.42 | 1.28 | 1.44 | 0.025 | **0.022** | 1.142 | 0.22 |

Comparing the TGB method with the FIA method, for 3D geometries, the limit value of $\phi_t$, over which the TGB method has a lower memory usage, is significantly higher. For 3D RAS geometries, the TGB method has a lower memory usage than the FIA for $\phi_t > 0.86$ ($\phi > 0.6$).

Unfortunately, the precise, general determination of limit $\phi_t$ is not possible because, in contrast to the CM, the memory overhead for the FIA depends on geometry porosity $\phi$. Since the memory overhead for the TGB method depends almost entirely on $\phi_t$, and given that, as we showed before, $\phi_t$ depends mainly on parameters other than $\phi$, for different geometries, the ratio of overheads for the TGB and FIA methods may be different. For 2D geometries, we may only estimate from Eqn. (31) that the FIA results in lower memory usage than the TGB method for $\phi_t \lesssim 0.5$.

Notice that the TGB method allows us to achieve a very high reduction in memory usage for all tested 2D geometries. For both ChipA_32 and ChipB_32 geometries, the total memory usage for the FIA/CM methods is $1.6 - 1.7\times$ higher than for the TGB method (2.28/1.43 and 2.44/1.43). Less favorable results are obtained for the D3Q19 lattice arrangement: the memory usage for the FIA/CM is only 10-33% (2.15/1.95 and 2.47/1.86) higher at most.

The memory overhead for the T2C method is always higher than for the TGB method. For the D3Q19 lattice arrangement, the T2C method requires only 12-14% more memory than the TGB method; but, for D2Q9, the T2C method requires about $1.7\times$ more memory than the TGB method. Such a large difference in memory usage for D2Q9 is caused by the very low $\Delta^M_{rc}$ for the TGB method, due to large tiles and a small number of ghost buffers.

The comparison between memory usage for the T2C and the CM methods shows that the limit value of $\phi_t$, over which T2C has lower memory usage, is significantly higher than for TGB and CM methods. For both 2D and 3D geometries, T2C will result in lower memory usage than the CM for $\phi_t \gtrsim 0.83$ ($\phi > 0.5$ for RAS). Compared with the FIA, T2C always involves higher memory usage, although differences for geometries with high $\phi_t$ are small.

For all 3D cases, and for the T2C method on the D2Q9 lattice, most of the tile memory overhead is $\Delta^M_{rc}$ (either the second copy of $f_i$ or ghost buffers). Only for the TGB method on the D2Q9 lattice is the memory overhead close to $\Delta^M_{\phi_t}$, which can be treated as an overhead for the "ideal" implementation of tiles. The removal of $\Delta^M_{rc}$ could significantly decrease memory requirements for tiles to values lower than for CM/FIA implementations.

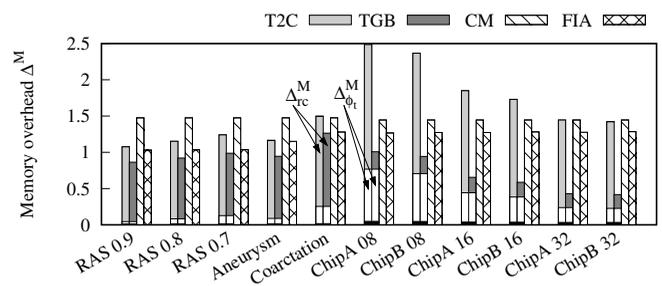

Fig. 8. Memory overhead estimates for tiles, CM and FIA implementations. Small black rectangles (almost invisible) denote $\Delta^M_{nt}$ and $\Delta^M_{ad}$.

### 4.1.3 Bandwidth overhead

The comparison of memory bandwidth overhead estimates for sparse test geometries is shown in Table 1 and Fig. 9. For all tested geometries, the overhead estimates for both versions of tile processing are significantly lower than for the CM/FIA (up to $10\times$ lower for the CM and up to $50\times$ lower than for the FIA). These results show that the tiles allow us to achieve the highest performance for sparse geometries. For the presented cases, better implementations for tiles could have involved performance that was only a few percent lower than the peak performance for dense geometries, which is much higher than for the CM technique, which could achieve at most $1/1.22 = 0.82$ of dense implementation performance due to $\Delta^B = 0.22$.

The bandwidth overhead for tiles is determined by two components: $\Delta^B_{nt}$ and $\Delta^B_{ad}$. In all cases, $\Delta^B_{ad}$ is smaller than $\Delta^B_{nt}$; thus, performance optimizations should focus on minimizing transfers of the node type fields. However, given that, for all test geometries, the overall bandwidth overhead is lower than 0.05, it can be difficult to decrease it further.

TABLE 2
Performance comparison for dense geometries. For [3], the collision/fluid model is deduced from information in the paper. Results presented in other work are sorted in descending order with respect to $BU$ for each combination of lattice arrangement and precision.

| work | model | MLUPS | $BU$ | lattice | precision | GPU |
|---|---|---|---|---|---|---|
| [10] | BGK q-compr. | 649 | 0.790 | D3Q19 | DP | Tesla K20x |
| [6] | BGK incompr. | 292 | 0.616 | D3Q19 | DP | Tesla C2070 |
| [14] | BGK q-compr. | 234 | 0.480 | D3Q19 | DP | Tesla M2050 |
| [1] | BGK q-compr. | 1036 | 0.757 | D3Q19 | SP | Tesla K20c |
| [4] | MRT incompr. | 516 | 0.701 | D3Q19 | SP | GTX 295 |
| [4] | BGK q-compr. | 512 | 0.696 | D3Q19 | SP | GTX 295 |
| [7] | BGK q-compr. | 400 | 0.543 | D3Q19 | SP | GTX 260 |
| [3] | BGK q-compr. | 300 | 0.528 | D3Q19 | SP | 8800 GTX |
| [9] | MRT incompr. | 443 | 0.455 | D3Q19 | SP | Tesla M2070 |
| [8] | BGK q-compr. | 375 | 0.321 | D3Q19 | SP | GTX 480 |
| [11] | BGK q-compr. | 874 | 0.656 | D2Q9 | SP | K5000m |
| [5] | BGK q-compr. | 947 | 0.481 | D2Q9 | SP | GTX 280 |
| [2] | BGK incompr. | 670 | 0.465 | D2Q9 | SP | 8800 Ultra |
| this | BGK incompr. | 682 | 0.719 | D3Q19 | DP | GTX Titan |
| this | BGK q-compr. | 639 | 0.674 | D3Q19 | DP | GTX Titan |
| this | MRT incompr. | 473 | 0.499 | D3Q19 | DP | GTX Titan |
| this | MRT q-compr. | 476 | 0.502 | D3Q19 | DP | GTX Titan |
| this | BGK incompr. | 1060 | 0.529 | D2Q9 | DP | GTX Titan |
| this | BGK q-compr. | 1020 | 0.509 | D2Q9 | DP | GTX Titan |
| this | MRT incompr. | 920 | 0.459 | D2Q9 | DP | GTX Titan |
| this | MRT q-compr. | 865 | 0.432 | D2Q9 | DP | GTX Titan |

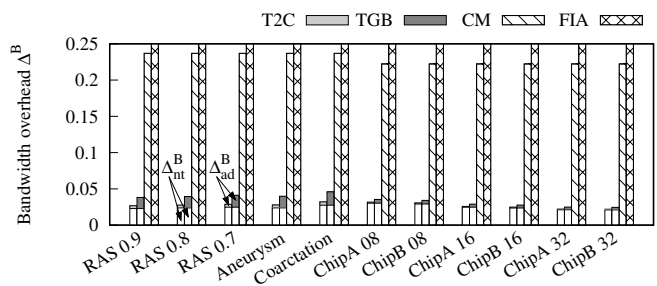

Fig. 9. Bandwidth overhead estimates for sparsity handling techniques. The overheads for the FIA exceed $1.0$ for all cases.

## 4.2 Performance comparison

To provide some measure that allows for a comparison of different implementations of lattice arrangements and on different machines, we calculated the number of bytes transferred per node from Eqn. (10) and the minimum required memory bandwidth to achieve the reported performance $P$ [MLUPS] as follows: $B_{case} = P \cdot B_{node}$. For sparse geometries, we only took into account the transfers for non-solid nodes. Next, we compared this value with the maximum theoretical GPU bandwidth $B_{GPU}$ and calculated a bandwidth utilization:

$$BU = \frac{B_{case}}{B_{GPU}} \qquad (43)$$

as shown in Table 2. This measure is based on [20] and similar to measures used in many papers (for example, [6], [7] and [11]). Due to its strict assumptions, it shows how the given implementation is close to the ideal for bandwidth-bound codes. According to the estimations presented in Section 2.2, the LBM implementations in GPUs are bandwidth-bound, except for either the implementations of complex models in previous generations of GPUs or lattices with a high number of links [30]. Notice also that we only used the theoretical maximum of memory bandwidth instead of the measured maximum.

### 4.2.1 Dense geometries

To illustrate how the proposed method has only small overheads, compared to the best implementations for dense geometries, Table 2 provides a performance comparison of different LBM implementations for dense geometries. We omitted results where performance was limited by the low computational power of the GPU, for example, double-precision computations on GTX280 [5], where $BU < 0.3$.

For the D3Q19 lattice arrangement and double precision, only [10] has $1.17\times$ higher $BU$ than the tile-based solution. Our implementation had $BU$ better than that reported by [6] ($1.17\times$) and [14] ($1.4\times$). Furthermore, many implementations for single-precision data had lower $BU$ than our version. Since we compared our solution, which was designed for sparse geometries, with highly optimized implementations for dense geometries (without overheads), these results show a low additional cost imposed by tiles.

The results for the D2Q9 lattice arrangement were less favorable for our implementation. Despite the higher ratio of bandwidth to computations (Section 2.2) and the lower register pressure, due to the size of data per node being about $2\times$ smaller, the tile-based implementation for D2Q9 had up to $1.36\times$ lower $BU$ than for the D3Q19 lattice. This low performance results from many synchronization points and a high complexity of code for ghost buffers handling. After removing the barriers, the performance for the D2Q9 lattice and the BGK incompressible model increased to 1,300 MLUPS ($BU = 0.649$). When we completely removed the code responsible for ghost buffers handling, the performance was 1,615 MLUPS ($BU = 0.806$). The second reason for a low performance is a lack of an optimized memory layout in kernels for the D2Q9 lattice, resulting in uncoalesced memory transactions.

Table 2 also highlights an another phenomenon of interest: although, in theory, the floating-point calculations should be completely masked by memory transfers [4], the performance of our code decreased in respect of more complex collision and fluid models. This results from explicit barriers in our code: after removing all synchronization points from the D2Q9-BGK incompressible kernel, the performance was the same for versions with and without collision computations. The decrease in performance for more complex models is also caused by a lower GPU occupancy resulting from increased register usage (especially visible for MRT on the D3Q19 lattice).

### 4.2.2 Sparse geometries

A performance comparison for sparse 3D geometries is shown in Table 3. For the CM method, we only present results from [18], which are for the optimized implementation of [15] in more recent hardware.

In all presented cases, $BU$ for the tile-based solution is significantly higher than was the case for [18] (up to $1.49\times$ in the aneurysm case). Compared with the FIA implementation in [19], $BU$ in our version is about $3\times$ better, despite a more complex lattice arrangement (D3Q19 vs. D3Q15). These results indicate that a tile-based solution allows us to achieve much better performance than indirect addressing methods for geometries with high tile porosity ($\phi_t > 0.8$ for all cases in Table 3).

Although, in the aneurysm case, we compared our single-GPU implementation with the multi-GPU version from [18], the overheads for handling multiple GPUs in [18] are low. For single-precision and large geometries, $BU$ for the multi-GPU version is usually $\sim 0.9$ of the single-GPU $BU$, e.g., 1,016 vs 288 MLUPS. Even if we assume that the single-GPU version of [18] achieves $1/0.8 = 1.25\times$ higher $BU$ (which is unlikely), the value of $BU$ for our implementation is still $1.19\times$ higher.

We have also observed that the measured performance is approximately proportional to $\phi_t$, which is consistent with estimations from Section 3.1.2: the estimated performance is proportional to $1/\left(1 + \Delta^B\right) = \phi_t/(\phi_t + constant)$, which is almost linear for large $\phi_t$. For example, for BGK quasi-compressible 2D and 3D kernels, $BU$ may be roughly estimated as $0.78 \cdot \phi_t - 0.13$, whereas, for BGK quasi-compressible 2D kernels only, the values of $BU$ are almost exactly equal to $0.56 \cdot \phi_t + 0.01$. Since, for all geometries used in [18] and [19], the value of $\phi_t$ is high, this may indicate that many real geometries allow us to achieve high tile porosity and high performance as a consequence.

### 4.2.3 Bandwidth overhead

In addition to direct performance measurements, we also measured the real bandwidth overhead of our implementation and compared it with the theoretical estimations in Section 3.1.2. The measured bandwidth overhead is defined as the number of additional 32-byte memory transactions relative to the minimal values resulting from Eqn. (10). For example, for the coarctation geometry with 659,105 non-solid nodes, the minimal bandwidth is $659105 * B_{node}/32 = 6261498$ transactions. Given that, in our implementation, we observed 6,733,121 transactions (3,401,855 reads and 3,331,266 writes after code profiling using *nvprof*), the measured overhead is $6733121/6261498 - 1 \approx 0.075$. The results are presented in Table 5.

The values in Table 5 show that the performance penalty for sparse geometries is higher than is apparent from Eqn. (35) and Eqn. (37). This results, inter alia, from additional memory traffic due to burst transactions being partially filled with data on non-solid nodes. The bandwidth overhead for tiles with low porosity is between values defined in Eqn. (35), Eqn. (37), Eqn. (41) and Eqn. (42). These observations show that the performance of our code could be improved by a better memory layout, which was attuned to porous tiles. However, since the real number of transactions for 3D geometries is only 2-8% larger than the minimal values, significant performance improvements may be difficult to achieve.

Notice that, for cavity3D geometry, the measured bandwidth overhead is even lower than the minimal estimated value $\Delta^B$. This results from very good cache utilization (we observed a cache hit ratio equal to 0.23), mainly for reads of node type values. On the other hand, the observed number of transactions for RAS_0.9 geometry is larger by 2.6 transactions per non-solid tile, compared with memory traffic resulting from $\Delta^B_{bt}$ (about 624 transactions per non-solid tile). This is caused by the uncoalesced and uncached transfers of additional tile data.

## 5 CONCLUSIONS

In this paper, GPU-based LBM implementation for fluid flows in sparse geometries was presented. In contrast to previous propositions for sparse geometries, which use indirect node addressing, our solution covers geometry with a uniform mesh of square tiles. Two implementations were presented: TGB and a single copy of the PDFs, and T2C of PDF values.

For the presented method, we have provided a detailed theoretical model, which allows us to analyze the memory and bandwidth overheads for different tile sizes, machine constraints (e.g., memory transaction size) and geometry layouts (defined, for example, by factors describing the porosity). The model has been used to compare the overheads introduced by tiles with the overheads of other sparse LBM implementations. It can also be applied to determine the "quality" of implementation and find code inefficiencies through the comparison of real performance with theoretical limits. We have also shown that real performance is approximately proportional, and that memory usage is inversely proportional, to a tile porosity.

For all the analyzed sparse geometries, the tile-based implementations resulted in the highest performance. For the T2C method, we achieved up to 682 MLUPS (71.9% utilization of maximum theoretical memory bandwidth) on the GTX Titan for the D3Q19 lattice, double-precision data and the BGK incompressible model. Based on theoretical estimates, for the presented cases, the T2C method required only 2-5% more data traffic than implementations for dense geometries. This is a significant reduction (up to $10\times$ more than for the CM and up to $50\times$ more than for the FIA) in the amount of ancillary data transfers required to handle geometry sparsity. Although real implementation leads to additional memory transfers, due to uncoalesced transactions, for the D3Q19 lattice arrangement, tiles with two copies allowed us to achieve up to $1.49\times$ higher device memory bandwidth utilization, compared to the CM technique (in the aneurysm case) and up to $3\times$ higher than for the FIA (in the coarctation case). The performance of our implementation was even close to the fastest implementations for dense geometries: we achieved 85% of bandwidth utilization, compared with [10].

TABLE 3
Performance comparison for sparse 3D geometries. The values in the first row are estimated, based on results presented in [19].

| Case | Model | this work | | other work | | | | | | |
|---|---|---|---|---|---|---|---|---|---|---|
| | | MLUPS | $BU$ | work | method | MLUPS | $BU$ | lattice | precision | GPU |
| Coarctation | BGK q-compr. | 574 | **0.605** | [19] | FIA | ∼ 150 | ∼ 0.2 | D3Q15 | DP | GTX 680 |
| Aneurysm | BGK q-compr. | 572 | **0.603** | [18] | CM | 1090 | 0.404 | D3Q19 | SP | 4 × Tesla C1060 |
| RAS_0.7 | BGK q-compr. | 565 | **0.596** | [18] | CM | 334 | 0.488 | D3Q19 | DP | Tesla K20 |
| RAS_0.8 | BGK q-compr. | 558 | **0.588** | [18] | CM | 330 | 0.482 | D3Q19 | DP | Tesla K20 |
| RAS_0.9 | BGK q-compr. | 558 | **0.588** | [18] | CM | 337 | 0.493 | D3Q19 | DP | Tesla K20 |

TABLE 4
Performance for sparse 2D geometries.

| Case | BGK incompr. | | BGK q-compr. | | MRT incompr. | | MRT q-compr. | |
|---|---|---|---|---|---|---|---|---|
| | MLUPS | $BU$ | MLUPS | $BU$ | MLUPS | $BU$ | MLUPS | $BU$ |
| ChipB_32 | 961 | 0.480 | 938 | 0.468 | 832 | 0.415 | 850 | 0.424 |
| ChipA_32 | 956 | 0.477 | 943 | 0.471 | 832 | 0.415 | 849 | 0.424 |
| ChipB_16 | 861 | 0.430 | 842 | 0.420 | 733 | 0.366 | 751 | 0.375 |
| ChipA_16 | 842 | 0.420 | 822 | 0.414 | 705 | 0.352 | 722 | 0.360 |
| ChipB_08 | 697 | 0.348 | 682 | 0.341 | 587 | 0.293 | 600 | 0.300 |
| ChipA_08 | 673 | 0.336 | 659 | 0.329 | 564 | 0.282 | 578 | 0.289 |

TABLE 5
Comparison of estimated and measured bandwidth overheads for 3D (top) and 2D (bottom) geometries.

| Case | $\Delta^B$ | $\Delta^B_{bt}$ | measured |
|---|---|---|---|
| cavity3D | 0.026 | 0.026 | 0.020 |
| RAS 0.9 | 0.027 | 0.058 | 0.062 |
| RAS 0.8 | 0.028 | 0.096 | 0.069 |
| RAS 0.7 | 0.029 | 0.141 | 0.079 |
| Aneurysm | 0.028 | 0.102 | 0.079 |
| Coarctation | 0.032 | 0.271 | 0.075 |
| ChipA 32 | 0.025 | 0.326 | 0.113 |
| ChipB 32 | 0.025 | 0.314 | 0.115 |
| ChipA 16 | 0.029 | 0.539 | 0.143 |
| ChipB 16 | 0.028 | 0.475 | 0.149 |
| ChipA 08 | 0.035 | 0.874 | 0.205 |
| ChipB 08 | 0.034 | 0.812 | 0.216 |

Tiles can also decrease memory usage, compared with FIA/CM implementations. For all tested geometries, the TGB method resulted in the lowest memory usage. Especially good results were observed for 2D cases with high tile porosity, where the estimated memory usage for FIA/CM techniques was up to $1.6 - 1.7\times$ higher than for tiles with ghost buffers. Combining this with very low bandwidth overheads, the tiles outperform FIA/CM methods when a geometry allows us to achieve high tile porosity.

The memory usage of the T2C method was between the FIA and the CM for all but one of the 3D geometries, as well as significantly higher than the FIA/CM for many 2D cases. Since the T2C method offers much a higher performance than the TGB method, the former is the preferred solution for 3D geometries, especially given that memory usage for the T2C method is often lower than for the CM, which is the fastest indirect addressing technique.

Since the gap between memory bandwidth and computational performance grows continuously, the methods decreasing bandwidth usage should gain in significance. The proposed technique is ideally suited to this trend and combines very high performance with low memory usage. We believe that it may also enhance other methods; for example, tiles can be used as a low-level data structure in multi-grid LBM implementations. Moreover, it should also be possible to adapt tiles to other stencil algorithms for sparse geometries, although additional research is needed. The wider application of the presented method requires a more generic tile-based library with parametrized code, which allows for different computational methods. Since tile porosity depends on unknown factors, additional studies of how the characteristics of sparse geometry impacts the performance of tiles may be necessary.

Our implementation achieves a level of performance close to the best dense ones. We hope that its future improvements, especially in memory usage, can make tiles a viable alternative for specialized implementations for dense geometries and remove a slightly artificial distinction between dense and sparse geometries. Since the global memory coalescing rules for Pascal and Volta GPUs are the same as for Kepler architecture used in this work, our implementation should achieve about 2 GLUPS per single Tesla V100 device, due to a higher memory bandwidth (900 GB/s) and better memory controllers.

The main drawbacks of the presented method are a rapid decrease in performance and an increase in memory usage when the tiles contain solid nodes, as well as a complex code, especially for tiles with ghost buffers, which requires many synchronization points that hinder performance.

Future work includes a multi-GPU version, which will be able to quickly simulate large geometries. We also intend to search for memory layouts, which are attuned to tiles with low porosity, and better tiling algorithms, which should allow us to achieve greater tile porosity.

## ACKNOWLEDGMENTS

The authors are very grateful to the reviewer's valuable comments, which improved the manuscript. This work was partially supported by the National Science Centre's Grant no. N N501 042140 and by the statutory funds belonging to the Chair of Computer Engineering, Faculty of Electronics, Wroclaw University of Science and Technology.

## REFERENCES

[1] M. J. Mawson and A. J. Revell, "Memory transfer optimization for a lattice Boltzmann solver on Kepler architecture nVidia GPUs," *Computer Physics Communications*, vol. 185, no. 10, pp. 2566–2574, 2014.

[2] J. Tölke, "Implementation of a Lattice Boltzmann kernel using the Compute Unified Device Architecture developed by nVIDIA," *Computing and Visualization in Science*, vol. 13, no. 1, p. 29, 2008.

[3] P. Bailey, J. Myre, S. Walsh, D. Lilja, and M. Saar, "Accelerating lattice Boltzmann fluid flow simulations using graphics processors," in *Parallel Processing, 2009. ICPP '09. International Conference on*, pp. 550–557.

[4] C. Obrecht, F. Kuznik, B. Tourancheau, and J.-J. Roux, "A new approach to the lattice Boltzmann method for graphics processing units," *Computers & Mathematics with Applications*, vol. 61, no. 12, pp. 3628–3638, 2011.

[5] F. Kuznik, C. Obrecht, G. Rusaouen, and J.-J. Roux, "LBM based flow simulation using GPU computing processor," *Computers & Mathematics with Applications*, vol. 59, no. 7, pp. 2380–2392, 2010.

[6] J. Habich, C. Feichtinger, H. Köstler, G. Hager, and G. Wellein, "Performance engineering for the lattice Boltzmann method on GPGPUs: Architectural requirements and performance results," *Computers & Fluids*, vol. 80, no. 0, pp. 276–282, 2013.

[7] P. Rinaldi, E. Dari, M. Vénere, and A. Clausse, "A lattice-Boltzmann solver for 3D fluid simulation on GPU," *Simulation Modelling Practice and Theory*, vol. 25, no. 0, pp. 163–171, 2012.

[8] M. Astorino, J. Sagredo, and A. Quarteroni, "A modular lattice Boltzmann solver for GPU computing processors," *SeMA Journal*, vol. 59, no. 1, pp. 53–78, 2012.

[9] C. Obrecht, F. Kuznik, B. Tourancheau, and J.-J. Roux, "Scalable lattice Boltzmann solvers for CUDA GPU clusters," *Parallel Computing*, vol. 39, no. 6–7, pp. 259–270, 2013.

[10] M. Januszewski and M. Kostur, "Sailfish: A flexible multi-GPU implementation of the lattice Boltzmann method," *Computer Physics Communications*, vol. 185, no. 9, pp. 2350–2368, 2014.

[11] M. J. Mawson, "Interactive fluid-structure interaction with many-core accelerators," Ph.D. dissertation, The University of Manchester, Manchester, UK, 2014.

[12] C. Feichtinger, J. Habich, H. Köstler, G. Hager, U. Rüde, and G. Wellein, "A flexible Patch-based lattice Boltzmann parallelization approach for heterogeneous GPU–CPU clusters," *Parallel Computing*, vol. 37, no. 9, pp. 536–549, 2011.

[13] C. Feichtinger, J. Habich, H. Köstler, U. Rüde, and T. Aoki, "Performance modeling and analysis of heterogeneous lattice Boltzmann simulations on CPU–GPU clusters," *Parallel Computing*, vol. 46, pp. 1–13, 2015.

[14] F. Schornbaum and U. Rüde, "Massively parallel algorithms for the lattice Boltzmann method on nonuniform grids," *SIAM Journal on Scientific Computing*, vol. 38, no. 2, pp. C96–C126, 2016.

[15] M. Bernaschi, M. Fatica, S. Melchionna, S. Succi, and E. Kaxiras, "A flexible high-performance lattice Boltzmann GPU code for the simulations of fluid flows in complex geometries," *Concurr. Comput. : Pract. Exper.*, vol. 22, no. 1, pp. 1–14, Jan. 2010.

[16] M. Bernaschi, S. Matsuoka, M. Bisson, M. Fatica, T. Endo, and S. Melchionna, "Petaflop biofluidics simulations on a two million-core system," in *2011 International Conference for High Performance Computing, Networking, Storage and Analysis (SC)*, pp. 1–12.

[17] M. Bernaschi, M. Bisson, M. Fatica, and S. Melchionna, "20 petaflops simulation of proteins suspensions in crowding conditions," in *2013 SC - International Conference for High Performance Computing, Networking, Storage and Analysis (SC)*, pp. 1–11.

[18] C. Huang, B. Shi, Z. Guo, and Z. Chai, "Multi-GPU based lattice Boltzmann method for hemodynamic simulation in patient-specific cerebral aneurysm," *Communications in Computational Physics*, vol. 17, no. 4, pp. 960–974, 2015.

[19] C. Nita, L. Itu, and C. Suciu, "GPU accelerated blood flow computation using the lattice Boltzmann method," in *High Performance Extreme Computing Conference (HPEC), 2013 IEEE*, pp. 1–6.

[20] G. Wellein, T. Zeiser, G. Hager, and S. Donath, "On the single processor performance of simple lattice Boltzmann kernels," *Comput. Fluids*, vol. 35, no. 8-9, pp. 910–919, Sep. 2006.

[21] G. Hager and G. Wellein, *Introduction to High Performance Computing for Scientists and Engineers*, 1st ed. Boca Raton, FL, USA: CRC Press, Inc., 2010.

[22] O. Filippova and D. Hänel, "Grid refinement for lattice-BGK models," *Journal of Computational Physics*, vol. 147, no. 1, pp. 219–228, 1998.

[23] M. Mazzeo and P. Coveney, "HemeLB: A high performance parallel lattice-Boltzmann code for large scale fluid flow in complex geometries," *Computer Physics Communications*, vol. 178, no. 12, pp. 894–914, 2008.

[24] D. Groen, J. Hetherington, H. B. Carver, R. W. Nash, M. O. Bernabeu, and P. V. Coveney, "Analysing and modelling the performance of the HemeLB lattice-Boltzmann simulation environment," *Journal of Computational Science*, vol. 4, no. 5, pp. 412–422, 2013.

[25] A. Parmigiani, C. Huber, O. Bachmann, and B. Chopard, "Pore-scale mass and reactant transport in multiphase porous media flows," *Journal of Fluid Mechanics*, vol. 686, p. 40–76, 2011.

[26] J. Fietz, M. J. Krause, C. Schulz, P. Sanders, and V. Heuveline, *Optimized Hybrid Parallel Lattice Boltzmann Fluid Flow Simulations on Complex Geometries*. Berlin, Heidelberg: Springer Berlin Heidelberg, 2012, pp. 818–829.

[27] M. Hasert, K. Masilamani, S. Zimny, H. Klimach, J. Qi, J. Bernsdorf, and S. Roller, "Complex fluid simulations with the parallel tree-based lattice Boltzmann solver Musubi," *Journal of Computational Science*, vol. 5, no. 5, pp. 784–794, 2014.

[28] R. D. M. Travasso, E. Corvera Poiré, M. Castro, J. C. Rodrguez-Manzaneque, and A. Hernández-Machado, "Tumor angiogenesis and vascular patterning: A mathematical model," *PLoS ONE*, vol. 6, no. 5, p. e19989, May 2011.

[29] K. Y. Lin, M. Maricevich, N. Bardeesy, R. Weissleder, and U. Mahmood, "In vivo quantitative microvasculature phenotype imaging of healthy and malignant tissues using a fiber-optic confocal laser microprobe," *Translational Oncology*, vol. 1, no. 2, p. 84–94, Jul. 2008.

[30] E. Calore, A. Gabbana, J. Kraus, S. F. Schifano, and R. Tripiccione, "Performance and portability of accelerated lattice Boltzmann applications with OpenACC," *Concurrency and Computation: Practice and Experience*, vol. 28, no. 12, pp. 3485–3502, 2016.

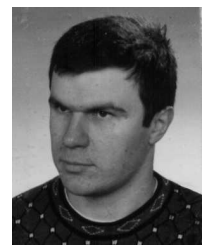

**Tadeusz Tomczak** received M.S. and Ph.D. degrees in computer science all from the Institute of Computers, Control, and Robotics, Wroclaw University of Technology, Poland, in 2002 and 2007 respectively. His research interests include residue number systems, fast computational hardware and parallel computing.

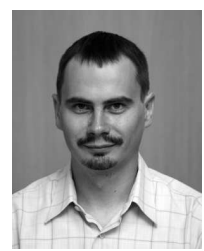

**Roman Grzegorz Szafran** received M.S. and Ph.D. degrees in chemical engineering all from the Faculty of Chemistry, Department of Chemical Engineering, Wroclaw University of Technology, Poland, in 1999 and 2004 respectively. His research interests include CFD/lattice-Boltzmann modeling of flow hydrodynamics in microfluidic devices, BIO-MEMS, multiphase flows, coating and encapsulation.